\documentclass[twocolumn,floatfix,showpacs,prl,aps,tightenlines,superscriptaddress]{revtex4}
\usepackage{graphicx}
\usepackage{psfrag}
\usepackage{dcolumn}
\usepackage{bm}

\newcommand{\lb}{{\cal L}_\beta}
\newcommand{\dt}{\partial_0}

\newcommand{\bea}{\begin{eqnarray}}
\newcommand{\eea}{\end{eqnarray}}
\newcommand{\beq}{\begin{equation}}
\newcommand{\eeq}{\end{equation}}

\begin{document}

\title{Accurate Evolutions of Orbiting Black-Hole Binaries Without Excision}

\author{M. Campanelli}  \affiliation{Department of Physics and Astronomy,
and Center for Gravitational Wave Astronomy,
The University of Texas at Brownsville, Brownsville, Texas 78520}

\author{C. O. Lousto} \affiliation{Department of Physics and Astronomy,
and Center for Gravitational Wave Astronomy,
The University of Texas at Brownsville, Brownsville, Texas 78520}

\author{P. Marronetti} \affiliation{Department of Physics,
Florida Atlantic University, Boca Raton, FL 33431}

\author{Y. Zlochower} \affiliation{Department of Physics and Astronomy,
and Center for Gravitational Wave Astronomy,
The University of Texas at Brownsville, Brownsville, Texas 78520}

\date{\today}

\begin{abstract}
We present a new algorithm for evolving orbiting black-hole binaries
that does not require excision or a corotating shift. Our algorithm is
based on a novel technique to handle the singular puncture conformal
factor.  This system, based on the BSSN formulation of Einstein's
equations, when used with a `pre-collapsed' initial lapse, is
non-singular at the start of the evolution, and remains non-singular
and stable provided that a good choice is made for the gauge. As a
test case, we use this technique to fully evolve orbiting black-hole
binaries from near the Innermost Stable Circular Orbit (ISCO) regime.
We show fourth order convergence of waveforms and compute the radiated
gravitational energy and angular momentum from the plunge. These results
are in good agreement with those predicted by the Lazarus approach.
\end{abstract}

 \maketitle

One of the most significant goals of numerical relativity is to
compute accurate gravitational waveforms from astrophysically
realistic simulations of merging black-hole binaries. The expectation of very
strong gravitational wave emission from the merger of two black holes,
and some of the newest astrophysical observations, from supermassive
galactic nuclei just about to merge
\cite{Komossa:2002tn} to stellar size black-hole binaries, make these 
systems one of the most extraordinary astrophysical objects under study today. 
Binary black hole mergers are expected not only
to provide information about the history and formation of the 
binary system but also to provide important precise tests of 
strong-field, highly dynamical relativity. 

Motivated by the forthcoming observations of ground-based gravitational wave
detectors, such as LIGO~\cite{LIGO}, and by
the next generation of space-based detectors, such as
LISA~\cite{Danzmann:2003tv}, the numerical relativity community
has dedicated a great deal of effort to solving the binary-black-hole
problem over the past few decades. After the `Binary Black Hole
Grand Challenge'~\cite{gc_web} several
new approaches have been pursued in the attempt to produce stable
three-dimensional (3D) numerical codes capable of evolving the full
Einstein field equations in the absence of any symmetry. This includes
the introduction of new formulations of these equations and the
development of numerical techniques for accurate evolutions of
black-hole binaries, such as higher order finite differencing,
spectral methods, and adaptive mesh refinement (AMR) (see
Ref. \cite{Alcubierre:2004es} and references therein).

The calculation of the gravitational radiation emitted from plunging
black-hole binaries was pioneered through the use of the Lazarus
approach, which bridges numerical relativity and perturbative
techniques to extract approximate gravitational waveforms
~\cite{Baker:2001sf, Baker:2002qf, Baker:2001nu}.
More recently important progress has been made toward evolving orbiting
binary-black-hole spacetimes with the use of stable full 3D numerical
relativity codes using corotating gauge conditions and
singularity excision~\cite{Bruegmann:2003aw,Alcubierre:2004hr,Pretorius:2005gq}.

Here we present a novel technique for evolving orbiting black holes
based on puncture data. This technique does not require a corotating
shift or singularity excision. Most importantly, we can produce
accurate complete waveforms from merging black-hole binaries.

In a previous paper we presented techniques for successfully
performing numerical relativity simulations of black-hole binaries with
fourth-order accuracy~\cite{Zlochower:2005bj}.  Our simulations are
based on a new coding framework, {\it LazEv}, which is built on top of the
Cactus Computational Toolkit~\cite{cactus_web}; this currently supports
higher order finite differencing for the BSSN formulation of
Einstein's equations~\cite{Nakamura87,Shibata95, Baumgarte99}, 
but is designed to be readily applicable to a broad class 
of formulations. Highly accurate
evolutions can be achieved using our unigrid higher-order
finite-difference code  along with a non-uniform coordinate system, such as
`Fisheye'~\cite{Baker:2001sf}, that concentrates gridpoints in the
central region containing the black holes.  

In the puncture approach~\cite{Brugmann:1997uc}
the metric on the initial slice is given
by~\cite{Brandt97b}
$
\gamma_{ab} = (\psi_{BL} + u)^4 \delta_{ab},
$
where
$
\psi_{BL} = 1 + \sum_{i=1}^n m_i/(2 r_i)
$
is the Brill-Lindquist conformal factor,
$m_i$ is the mass parameter of puncture $i$, $r_i$ is the coordinate
distance to puncture $i$, and $u$ is finite on the punctures. If the puncture
positions are fixed throughout the evolution, then the singular behavior in
the metric is contained in $\psi_{BL}$ and can thus be treated analytically.
However, holding the puncture fixed throughout the evolution leads to
significant coordinate distortions that tend to kill the run before a common
horizon forms.

We propose a method for evolving puncture type data without fixing the
puncture positions during the evolution. Our method is based on the
BSSN formulation. In the BSSN system one evolves
a conformal metric $\tilde \gamma_{ab} = \exp(-4 \phi) \gamma_{ab}$ which has unit 
determinant, $K=K^a_{\;\;a}$, the conformal trace-free extrinsic curvature 
$\tilde A_{ab} = \exp(-4 \phi) (K_{ab} - \gamma_{ab} K/3)$,
 the conformal exponent $\phi$, and $\tilde \Gamma^i = -\partial_j \tilde \gamma^{ij}$.
 In order to regularize the system near the puncture, 
we replace the BSSN exponent $\phi$ (which has an $O(\ln r)$
singularity at the puncture) with a new variable $\chi = \exp(-4
\phi)$ that is $C^4$ on the puncture. Additionally we modify the
standard 1+log lapse and Gamma-driver shift gauge
conditions~\cite{Alcubierre02a}. Our system is explicitly finite on
the initial slice. The evolution equations for this system are
\cite{Alcubierre02a}:
\begin{eqnarray}
   \label{eq:gt_evol}
   \dt \tilde \gamma_{ij} &=& -2 \alpha \tilde A_{ij}, \\
   \partial_t \chi &=& \frac{2}{3}\chi \left(\alpha K -  
     \partial_a \beta^a\right) + \beta^i \partial_i \chi, \label{eq:chi_evol}\\
   \dt \tilde A_{ij} &=& \chi \left(-D_i D_j \alpha +
       \alpha R_{ij}\right)^{TF} +\nonumber \\
        &\,&\alpha \left(K \tilde A_{ij} -
        2 \tilde A_{ik} \tilde A^{k}_{\,j}\right), \label{eq:at_evol}\\
   \dt K &=& - D^i D_i \alpha + \alpha \left(\tilde A_{ij}\tilde
         A^{ij} +\frac{1}{3}K^2\right),\\
  \partial_t \tilde \Gamma^i &=& \tilde \gamma^{jk} \partial_j
\partial_k \beta^i + \frac{1}{3} \tilde \gamma^{ij} \partial_j
\partial_k \beta^k + \beta^j \partial_j \tilde \Gamma^i - \nonumber \\
  &\,&\tilde
\Gamma^j \partial_j \beta^i + 
 \frac{2}{3}\tilde \Gamma^i \partial_j
\beta^j - 2 \tilde A^{i j}\partial_j \alpha + \nonumber \\
 &\,& 2 \alpha \left(\tilde
{\Gamma^i}_{jk} \tilde A^{jk} + 6 \tilde A^{ij}\partial_j \phi -
\frac{2}{3} \tilde \gamma^{ij} \partial_j K\right), \hspace{10mm}  \label{eq:dtGamma}
\end{eqnarray}
where $\dt = \partial_t - \lb$, $TF$ indicates that only the trace-free 
part of the tensor is
used, $R_{ij} = \tilde R_{ij} + {R^\phi}_{ij}$ is given by
\begin{eqnarray}
  {R^\phi}_{ij} &=& - 2 \tilde D_i \tilde D_j \phi - 2 \tilde
\gamma_{ij} \tilde D^k \tilde D_k \phi + 4 \tilde D_i \phi \tilde D_j \phi
- \nonumber \\
  &\,&4\tilde \gamma_{i j} \tilde D^k \phi \tilde D_k \phi, \\
 \label{eq:RT}
 \tilde R_{ij} &=& -\frac{1}{2} \tilde \gamma^{lm}\partial_l
\partial_m \tilde \gamma_{ij} + \tilde \gamma_{k(i}\partial_{j)}
\tilde \Gamma^k + \tilde \Gamma^k \tilde \Gamma_{(ij)k} + \nonumber \\
 &\,&\tilde
\gamma^{lm}\left(2 \tilde {\Gamma^k}_{l(i}\tilde \Gamma_{j)km} +
\tilde {\Gamma^k}_{im}\tilde \Gamma_{klj}\right),
\end{eqnarray}
$D_i$ is the covariant derivative with respect to $\gamma_{ij}$, and
$\tilde D_i$ is the covariant derivative with respect to $\tilde
\gamma_{ij}$.
$\tilde \Gamma^i$ is replaced by $-\partial_j \tilde
\gamma^{ij}$ in Eqs.~(\ref{eq:gt_evol}) - (\ref{eq:RT})
wherever it is not differentiated. Note that
Eqs.~(\ref{eq:chi_evol})~and~(\ref{eq:dtGamma}) give the $\partial_t$
derivatives of $\chi$ and $\tilde \Gamma^i$ rather than the
$\partial_0$ derivatives. The Lie derivatives of the non-tensorial
quantities ($\tilde \gamma_{ij}$, and $\tilde A_{ij}$) are given by
\begin{eqnarray}
  \lb \tilde \gamma_{ij} &=& \beta^k\partial_k \tilde \gamma_{ij} +
   \tilde \gamma_{ik} \partial_j \beta^k + \nonumber \\
     &\,&
    \tilde \gamma_{jk} \partial_i \beta^k -
    \frac{2}{3} \tilde \gamma_{ij} \partial_k \beta^k, \\
  \lb \tilde A_{ij} &=& \beta^k\partial_k \tilde A_{ij} +
   \tilde A_{ik} \partial_j \beta^k + \nonumber \\
     &\,&
    \tilde A_{jk} \partial_i \beta^k -
    \frac{2}{3} \tilde A_{ij} \partial_k \beta^k.
\end{eqnarray}
Note that
$
\partial_i \phi = -1 /(4 \chi) \partial_i \chi
$
and
$\partial_{ij} \phi = \frac{1}{4}(-\partial_{ij}\chi /\chi + 
                      \partial_i \chi \partial_j\chi / \chi^2)
$
are singular on the puncture; as a result, we identify several potentially singular
terms in Eqs.~(\ref{eq:gt_evol})-(\ref{eq:dtGamma}).
In Eq.~(\ref{eq:at_evol}) 
the term $D_i D_j \alpha$ can be expressed in terms of 
$\partial_i \phi \partial_j \alpha$ + non-singular terms. However this term 
is multiplied by $\chi$ and the product is $C^3$ on the puncture. Additionally, 
the $\partial_i\partial_j \phi$ terms (from $R_{ij}$) are multiplied by
$\alpha\chi$, and are thus $C^2$ on the puncture.
If the lapse $\alpha \sim r^2$ on the puncture (as is our choice)
then $\tilde A_{ab}$ is $C^4$ on the puncture (provided that it is $C^4$ on
the initial slice). However, in
Eq.~(\ref{eq:dtGamma}) we find the singular term $\alpha \tilde A^{ij}
 \partial_i \phi$. With our choice of initial data, $\tilde A^{ij} \sim r^2$, and thus
$\tilde \Gamma^i$ is $C^0$ on the puncture. We can then choose an
initial lapse $\alpha = \psi_{BL}^{-2}$ which is $O(r^2)$ on the
puncture.  With this choice of lapse, $\tilde \Gamma^i$ evolves to a
function that is $C^2$ on the puncture. However, $\tilde \Gamma^i$ will only
remain well behaved if the lapse condition maintains this $O(r^2)$ behavior
near the puncture. We found that the following gauge conditions
produced smooth waveforms:
\begin{eqnarray}
  \partial_0 \alpha &=& - 2 \alpha K, \\
  \partial_t \beta^a &=& B^a,\quad \partial_t B^a = 3/4 \partial_t \tilde \Gamma^a - \eta B^a.
\end{eqnarray}

In black-hole binary systems the features that need to be resolved range in
scale from a fraction of $M$ near the horizons to over $100M$
in the wave zone when black holes are in a slow inspiral motion. This
difference in scales makes simple unigrid evolutions extremely
inefficient. We mitigate this problem by introducing a `multiple
transition' Fisheye transformation. This Fisheye coordinate is a
natural extension of the `transition Fisheye'
coordinate~\cite{Baker:2001sf,Alcubierre02a} (where the effective
resolution changes from some inner resolution $h$ to an outer
resolution $a\, h$ inside a region of a given width, where $a$ is a
user specified parameter), but with multiple transition regions
allowing for fine tuning of the resolution in intermediate
regions. This new Fisheye coordinate simulates fixed-mesh refinement.

For puncture data the estimated ISCO~\cite{Baumgarte00a} is characterized
 by the parameters 
\begin{eqnarray}
&&L/M=4.9,\quad P/M=0.335,\quad Y/M=\pm1.1515,\nonumber \\
&&J/M^2=0.77,\quad M\Omega=0.178,\quad m=0.45M,
\end{eqnarray}
where $m$ is the mass of each single black hole, $M$ is the
total ADM mass of the binary system, $L$ is the proper distance
between the apparent horizons, $P$ is the magnitude of the linear
momenta (equal but opposite and perpendicular to the line connecting
the holes), $J$ is the total angular momentum, and $(0,Y,0)$ is the coordinate
location of the punctures.
We use the Brandt-Br\"ugmann approach along with the 
BAM\_Elliptic~\cite{Brandt97b,cactus_web} Cactus thorn to solve 
for these initial data.

We use $\pi$-rotational symmetry about the polar axis (Pi-symmetry)
and reflection symmetry across the orbital plane to reduce the
computational domain to one quadrant. We choose Fisheye parameters
that produce an inner resolution $h$, an
intermediate resolution of $6h$, and an outer resolution of $25 h$ (to
push the boundaries out to $60M$).

\begin{figure}
\begin{center}
\includegraphics[width=2.5in]{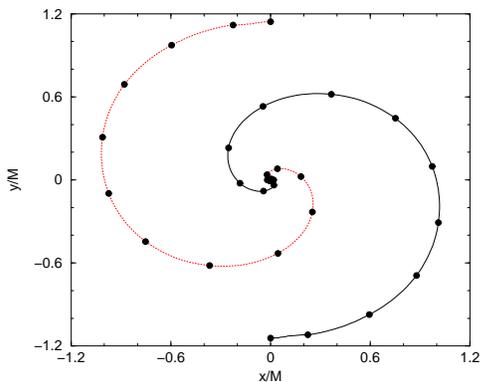}
\caption{The trajectories of the punctures (the circles correspond to positions 
at t=0, 2.5M, 5M, etc).
 The common horizon forms just after the puncture complete a half orbit. The punctures continue to
orbit throughout the evolution.}
\label{fig:ptrack}
\end{center}
\end{figure}

At the puncture $\chi=0$ and Eq.~(\ref{eq:chi_evol}) implies that 
the puncture position obeys $\partial_t {\vec x_{punct}} =
 -{\vec \beta} (\vec x_{punct})$. We use this to track the puncture positions
throughout the evolution.
Figure~\ref{fig:ptrack} shows the trajectory  of the punctures for an 
$h=M/24$ run. A common horizon forms just after the punctures complete half of
an orbit. The punctures continue to orbit throughout the evolution.

\begin{figure}
\begin{center}
\includegraphics[width=2.7in]{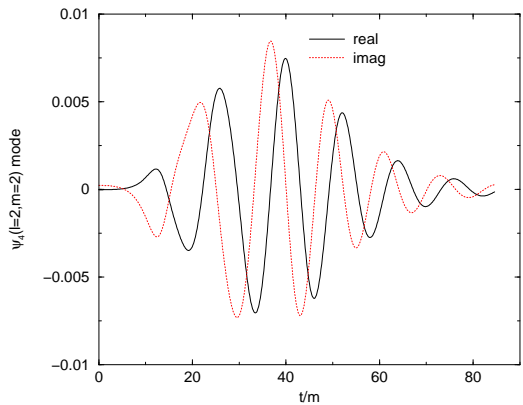}
\caption{QC0 waveform. The real and imaginary parts of the
$(\ell=2, m=2)$ mode of $\psi_4$ calculated at $r=10M$.}
\label{fig:whole_wave}
\end{center}
\end{figure}

\begin{figure}
\begin{center}
\includegraphics[width=2.7in]{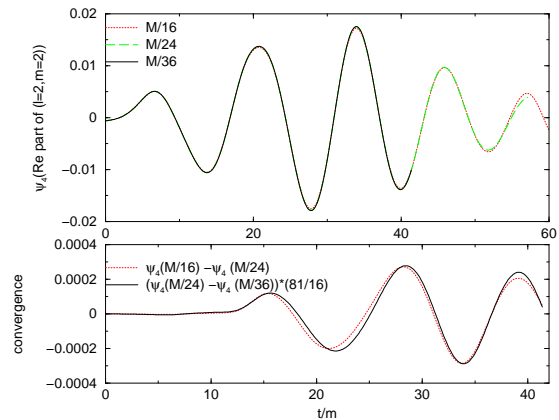}
\caption{QC0 waveforms. The top plot shows the real part of the $(\ell=2,
m=2)$ mode at $r=5M$ for resolutions of $h=M/16, M/24, M/36$. The bottom
plot shows the differences between waveforms for $h=M/16$ and $h=M/24$
as well as the difference between waveforms for $h=M/24$ and $h=M/36$.
The latter difference has been rescaled by $(3/2)^4$ to demonstrate fourth-order 
convergence. }
\label{fig:QC0Wfs}
\end{center}
\end{figure}

We use the Zorro thorn~\cite{Baker:2001sf,Zlochower:2005bj} to
calculate $\psi_4$ and decompose it into $(\ell, m)$ modes.
Figure~\ref{fig:whole_wave} shows the real and imaginary parts of the
$(\ell=2, m=0)$ mode at $r=10M$. 
We stopped this run at $\sim85M$ because of limited computational resources.
Figure~\ref{fig:QC0Wfs} shows the real part of the $(\ell=2,m=2)$ mode
of $\psi_4$ at $r=5M$ (we choose this small observer radius to delay outer
boundary effects) for resolutions of $h=M/16, M/24, M/36$, as well as
a convergence plot of these data.  The convergence plot shows that the
waveforms are fourth-order convergent.  The radiated
energy (as measured from the $M/24$ run) is $2.8\%\pm0.2\%$ in excellent
agreement with the final horizon mass (see below), and the radiated
angular momentum is $J_z = -0.11M^2\pm0.01 M^2$ (computed using Eqs. (22)-(24)
of Ref.~\cite{Campanelli99}).

We use Jonathan Thornburg's AHFinderDirect 
thorn~\cite{Thornburg2003:AH-finding} to find apparent horizons. We
first detect a common apparent horizon at  $t=19M$. The common 
horizon has an irreducible
mass of $0.9055M$ and the ratio of polar to equatorial circumferences
asymptotes to $0.898\pm0.002$.
One can show analytically \cite{Alcubierre:2004hr} that for a Kerr
black hole, the ratio of the polar and equatorial horizon
circumferences $C_r=C_p/C_e$
is given by
\begin{equation}
C_r = \frac{1+\sqrt{1-\tilde a^2}}{\pi} \, 
        E \left( -\frac{\tilde a^2}{(1+\sqrt{1-\tilde a^2})^2} \right),
\label{eq:Cr-exact}
\end{equation}
where $\tilde a = a/M_{\cal H}$ and $E(x)$ is the complete elliptic 
integral of the second kind. In the case of a perturbed black hole
 produced by a merger,
this ratio shows quasinormal ringing behavior before damping to the expected
Kerr value.
The horizon mass is related to the spin and irreducible mass by
$
M_{\cal H} = (M_{irr}/\tilde a)\sqrt{2 (1 - \sqrt{1-\tilde a^2})}.
$
Hence, the irreducible horizon mass and circumference ratio that we measure 
correspond to a spin of
$\tilde a = 0.683\pm0.006$ and a horizon mass of $0.973M\pm0.002M$. The
 horizon mass reduction is in 
excellent agreement with the calculated radiated energy of $0.028M\pm0.002M$.

Table \ref{table:results} summarizes the main results of our full numerical evolution
of binary black holes from the ISCO down to the final Kerr black hole remnant.
Waveforms, as described in terms of the Weyl scalar $\psi_4$, are dominated
by the modes $\ell=2, m=\pm2$ and show a strong circular polarization as seen
along the axis of orbital symmetry. These results are in good agreement 
with the results calculated by the Lazarus approach \cite{Baker:2001nu,Baker:2002qf}.

The Hamiltonian constraint violation along the $y$ axis (when the punctures are no
longer on the axis) shows cubic convergence  in the former
positions of the punctures and quadratic convergence outside. This quadratic
convergence is most likely due to our second-order-accurate initial data.
Further study is needed to see if constraint violations on the puncture
converge to zero.

\begin{table}
\caption{Results of the evolution.}
\begin{ruledtabular}
\begin{tabular}{lllll}\label{table:results}
Method & $E_{rad}/M$ & $J_{rad}/M^2$ & Merger time/M & $a/M_{\cal H}$ \\
\hline
This paper & $2.8\pm0.2$ & $14\pm1\%$ & $T_{CAH}\approx19$ & $0.683\pm0.006$ \\ 
Lazarus\protect\footnote{Errors quoted in the Lazarus runs
are only those from the differences among transition times; hence they 
only represent a lower bound to the total errors.} &
$2.5\pm0.2$ & $13\pm2\% $ & $T_{Tran}\approx10$ & $0.70\pm0.02$ \\
\end{tabular}
\end{ruledtabular}
\end{table}

Note that in this paper, we used a convenient numerical tetrad 
to calculate the Weyl scalar $\psi_4$. This waveform extraction procedure
is valid when the far-field spacetime approaches that of a 
perturbed Schwarzschild black hole. In future simulations 
we plan to use a more robust wave extraction 
method recently implemented in Ref. \cite{Campanelli:2005ia}.
This method is based on the use of a quasi-Kinnersley tetrad 
to compute $\psi_4$ and should allow us to extract waveforms closer to the hole,where outer boundary effects are less important.

The full nonlinear numerical technique just described has shown long-term
 stability, at least for the merger of binary black hole cases we have had
the opportunity to analyze so far.  We have also been able to prove
fourth-order convergence up to relatively high resolutions such as $h=M/36$.
This opens up the possibility of studying even more interesting
astrophysical scenarios such as black-hole binaries starting from
larger separations, which would undergo several orbits before
the final plunge. Additionally, we plan evolutions based on the more
astrophysically relevant Thin-Sandwich~\cite{Hannam:2005rp} and
Post-Newtonian~\cite{Tichy02} initial data sets. 
We will also consider unequal-mass black-hole binaries  and
compute their gravitational kick in order to evaluate its astrophysical
consequences~\cite{Campanelli:2004zw}. Finally, we plan to examine mergers
of highly spinning black-hole binaries  and study the possible `hang-up' 
of the binary until the excess angular momentum is radiated.

\acknowledgments
This work was recently presented at the ``Numerical Relativity 2005:
Compact
Binaries''~\cite{YZ_numrel2005_talk}. See~\cite{Choi_numrel2005_talk}
for a similar approach with comparable results.
 
We thank Erik Schnetter for providing the thorns to implement 
Pi-symmetry boundary conditions. We thank Bernard Kelly for careful reading
of this paper. We thank 
Mark Hannam for helpful discussions.
We gratefully acknowledge the support of the
NASA Center for Gravitational Wave Astronomy at University of Texas at
Brownsville (NAG5-13396) and the NSF for financial
support from grants PHY-0140326 and PHY-0354867. Computational resources
were provided by the Funes cluster at UTB. 
	
\bibliographystyle{apsrev}
\bibliography{../Lazarus/bibtex/references}
\thebibliography{letter1}

\end{document}